\documentclass[modern]{aastex62}
\usepackage{amsmath}
\graphicspath{{./}{figures/}}

\shorttitle{SPEs on Mars}
\shortauthors{Atri et al.}

\begin{document}

\title{Modeling Solar Proton Event-induced Martian Surface Radiation Dose}

\correspondingauthor{Dimitra Atri}
\email{atri@nyu.edu}

\author{Dimitra Atri}
\affiliation{Center for Space Science \\
New York University Abu Dhabi \\
PO Box 129188, Saadiyat Island, Abu Dhabi, UAE}
\affiliation{Blue Marble Space Institute of Science\\
1001 4th Ave, Suite 3201\\
Seattle, WA 98154, USA}
\author{Caitlin MacArthur}
\affiliation{Victoria University of Wellington\\Kelburn, Wellington 6012, New Zealand}
\affiliation{Blue Marble Space Institute of Science\\
1001 4th Ave, Suite 3201\\
Seattle, WA 98154, USA}
\author{Ian Dobbs-Dixon}
\affiliation{ New York University Abu Dhabi \\ PO Box 129188, Saadiyat Island, Abu Dhabi, UAE}
\affiliation{Center for Space Science \\
New York University Abu Dhabi \\
PO Box 129188, Saadiyat Island, Abu Dhabi, UAE}
\affiliation{Center for Astro, Particle and Planetary Physics (CAP3) \\ New York University Abu Dhabi}



\begin{abstract}

Solar Proton Events (SPEs) can cause abrupt and significant enhancements to the Martian surface radiation dose. Observations of the impact of SPEs on the Martian surface are available from satellites and surface detectors, but the data set is very limited ($\sim$20 years) in time, and the energy range is limited in scope, which makes it insufficient to estimate the impact of major events on the Martian surface. On the other hand, long-term data of SPEs impacting the Earth spanning a large energy range is widely available, and can be used to estimate the impact of major events on Mars on long timescales. Herein, we take major SPEs observed during the past several decades on Earth (1956 - 2014), along with PAMELA observations (2006 - 2014) and use the GEANT4 Monte Carlo code to calculate the Martian surface radiation dose. We study the contribution of proton fluence and spectral shape of events on the surface radiation dose and estimated the impact of possible major SPEs on the Martian surface in the future. These results have major implications for the planned human exploration of Mars. Overall we find that the radiation dose from extreme events can have a significant impact on astronaut health, and in rare, worst case scenarios, the estimated dose can even reach lethal levels.

\end{abstract}

\keywords{solar flares --- mars --- radiation dose}

\section{Introduction} \label{sec:intro}
Space weather impacts the Martian atmosphere by altering its chemistry and causing erosion \citep{jakosky2015maven, atri2016did}. Space weather-induced radiation exposure to satellites and potentially to astronauts is of great significance to space exploration, especially the planned human exploration of Mars \citep{cucinotta2001space}. Over the past few years, there has been an increased focus on the possibility of crewed missions to Mars, and one of the major obstacles in such missions is the adverse effects of ionizing radiation (charged particles) on the health of astronauts \citep{hellweg2007getting}. Unlike Earth, Mars has a thin atmosphere (about 2\% column density compared to the Earth) and no intrinsic global magnetic field to protect it from charged particles. A combination of these two factors exposes the Martian surface to a high radiation dose from a background flux of Galactic Cosmic Rays (GCRs) \citep{saganti2004radiation, gronoff2015computation} and episodic solar events \citep{townsend2011estimates}. Understanding the long-term Martian radiation environment in the context of space weather events is therefore vital in planning human exploration activities on the planet. 

While the solar extreme ultraviolet (XUV) emission can be easily blocked by a thin protective sheet, energetic charged particles ($\geq$ 1 GeV) have the capability to penetrate a shield, undergo hadronic interactions, and produce secondary particles such as electrons, muons, and neutrons. Among all the secondary particles produced from these interactions, neutrons are some of the most dangerous to human health \citep{mountford1992recommendations}. Secondary particles, depending on their dose, can cause irreparable biological damage to astronauts and also damage spacecraft electronics. It is therefore important to model the impact of space weather events on Martian surface radiation dose and their implications for human space exploration. One of the most important instruments for this purpose is the Radiation Assessment Detector (RAD) \cite{hassler2012radiation} on board Mars Science Laboratory (MSL) Curiosity rover, which has been measuring radiation doses on the Martian surface since 2012. Although there have not been many major Solar Proton Events (SPEs) since RAD was commissioned due to low solar activity, its measurements of background radiation dose from GCRs is highly valuable. Calculation of GCR-induced radiation dose corresponding to the measured GCR flux and the measured dose can be used to validate numerical models, which can then be used to accurately model the impact of SPEs on the Martian surface. 

There are observations of several minor SPEs over the past decade from the Martian orbit and surface \citep{larson2015maven, guo2015modeling}. This data has been valuable in understanding the impact of SPEs on the Martian surface and has been used for a number of important studies; for example \cite{matthia2016martian} used GEANT4 to model radiation dose from the background flux of GCRs and successfully matched it to RAD measurements. \cite{guo2018generalized} used 20 years of satellite data (SOHO/EPIN) of terrestrial SPEs and modeled the Martian surface radiation dose. While this approach is useful to estimate the surface radiation environment for minor events, its major disadvantage is the lack of observations of major events which usually occur on longer timescales. This smaller time window does not capture the full breadth of events, in terms of intensity, duration, and particle spectra. In addition, the energy spectra on these satellites is measured across a significantly limited energy range (eg. EPHIN measures protons between 4 - 50 MeV) and does not cover the full spectrum of charged particles estimated to be emitted in SPEs (10 MeV - 10 GeV) \citep{usoskin2011ionization}. On the other hand, major events have been studied in great detail on Earth for the past several decades. Data from terrestrial Ground Level Enhancements (GLEs) is available for longer timescales and also spans a much wider energy range, covering the entire spectrum of the event, unlike satellite data \citep{tylka2009new}. In this work we make use of terrestrial and satellite data of SPEs from 1956 to 2014, and model their interaction with the Martian atmosphere.

In the next section, we describe our numerical modeling methodology, details of code validation, and surface radiation dose calculations. In the results section, we discuss how we studied the variation in event spectra by normalizing the event fluence. We then use a scaling relation to estimate the radiation dose in case of a Carrington-like events and superflares. In the conclusions we discuss how enhanced radiation exposure from SPEs impacts astronaut health in future crewed missions to Mars. 

\section{Numerical Modeling} \label{sec:method}
We use the spectra measured with the Payload for Antimatter Matter Exploration and Light-nuclei Astrophysics (PAMELA) mission \citep{bruno2018solar}, estimated spectra from historical SPEs \citep{tylka2009new}, and one event measured with the Radiation Assessment Detector (September 2017) for our calculations. 30 events measured by PAMELA between July 2006 and September 2014 have been analyzed and presented in parametric form with protons in energies ranging from 80 MeV to 3 GeV in \cite{bruno2018solar}. The power law spectrum is given by:
\begin{equation}
    \Phi_{sep}(E) = A \times (E/E_{S})^{-\gamma} \times e^{-E/E_{0}},
\end{equation}
where $\gamma$ is the spectral index, the scaling energy E$_{S}$ is set to 80 MeV, the PAMELA energy threshold, E$_{0}$ is the rollover energy, and A is for normalization. The proton spectra of historical SPEs from 1956 to 2012 have been modeled in parametrized form by \cite{tylka2009new}. The event-averaged spectra are represented in the form of Band functions, with protons in energies ranging from 10 MeV to 10 GeV. The event integrated fluence, J (protons/cm$^{2}$) is given by \citep{usoskin2011ionization}:

\begin{equation}
\begin{split}
&J(>R) = J_{0} \times R^{-\gamma_{1}}e^{-R/R_{0}}, \hspace{0.2cm} \hspace{0.1cm} R \leq(\gamma_{2} - \gamma_{1})R_{0}, \\
    &J(>R) = J_{0} \times B \times R^{-\gamma_{2}},\hspace{0.2cm} \hspace{0.1cm} R > (\gamma_{2} - \gamma_{1})R_{0}, 
    \end{split}
\end{equation}
\begin{equation}
\begin{split}
& B = [(\gamma_{2}-\gamma_{1})R_{0}]^{(\gamma_{2}-\gamma_{1})}\times e^{(\gamma_{1}-\gamma_{2})}, \\&R = \sqrt{T^{2} + 2 T_{0} T}
\end{split}
\end{equation}
   
where R is the rigidity in GV, T is the kinetic energy in GeV, J$_{0}$ is for normalization and T$_{0}$ is the rest mass energy of the proton, 0.938 GeV. The values of parameters J$_{0}$, R$_{0}$, $\gamma_{1}$ and $\gamma_{2}$ are obtained from \cite{tylka2009new}, which are based on fits to the SPE data. Events with delayed particle enhancements, and those produced with other mechanisms, such as Energetic Storm Particles (ESPs) were discarded, and we used the remaining 59 events out of 70 for this work, which were all prompt events. Prompt events are all produced by the same mechanism, where particles are accelerated in the shock and travel directly along field lines. There are two common events in the two samples, December 13 2006 and May 17 2012.

We note that the same event observed on Earth will produce a different spectrum on Mars due to transport in the interplanetary medium, which depends on a number of factors (solar wind, interplanetary magnetic field, etc). However, for the purpose of this manuscript we assume that the spectral shape does not change, and thus we use the Earth spectra, scaled to the Martian orbit for our calculations. Since charged particle transport in the interplanetary medium is a complex process, requiring data on variations in magnetic fields, density etc., which is not available for most historical events, this was the most reasonable approach. A similar approach was followed by \cite{guo2018generalized} but for a limited number of events, and in a shorter energy range, and represents the upper limit of radiation exposure. Also vital for this calculation is a model of the Martian atmosphere, which we obtained from the Mars Climate Database (MCD) \citep{forget1999improved, millour2015mars}. The model includes a number of important quantities such as temperature, pressure, density, and atmospheric composition. We used the general atmospheric model based at the mean surface level for our simulations. For validation purposes, we used the atmospheric model located at the Gale Crater (location of RAD), which is at a depth of about 4 km below the mean surface level with a column depth of 22 g cm$^{-2}$. 
 
We used the GEANT4 numerical model to simulate particle propagation in the Martian atmosphere \citep{agostinelli2003geant4}. It is a widely used Monte Carlo model for modelling charged particle propagation in a variety of media in high-energy physics, medical, planetary and space sciences. The model simulates charged particle interactions with matter, including hadronic and electromagnetic interactions, scattering, and particle decays. Due to the popularity of the code, it is calibrated with numerous experiments around the globe, making it highly reliable for calculations. We have used GEANT4 in the past for similar computations to model charged particle-induced surface radiation dose on terrestrial exoplanets \citep{atri2013galactic, atri2016possibility,atri2020stellar} and subsurface radiation dose on Mars \citep{atri2016modelling}. As mentioned earlier, since \cite{matthia2016martian} have demonstrated the accuracy of GEANT4 for modeling of GCR-induced radiation dose by comparing it with RAD measurements \citep{hassler2014mars}, it was natural for us to choose their model configuration for our simulations. The physics lists we used were emstandard-opt3 and G4HadronPhysicsQGSP-BERT-HP, and the models were BertiniCascade, QGSP and FTFP for protons, and NeutronHPInelastic, BertiniCascade, QGSP and FTFP for neutrons. We validated our method by computing the GCR-induced background radiation dose at Gale Crater, which has been measured by RAD. We used the BON10 model \citep{o2010badhwar} to obtain the background GCR spectrum, with 87\% protons, 12\% alpha particles and 1\% Iron (as a substitute for heavier particles). We show the results in the next section. We have also made comparison of our SPE results with similar work on SPEs, which is also discussed in the next section.

For each SPE we simulated the atmospheric interaction of 10$^{9}$ primary protons in the 10 MeV - 10 GeV energy range. Due to lack of a magnetic field, particles were incident randomly from all over the hemisphere for each case. The energy spectrum of secondary particles was obtained on the surface from GEANT4 for each case. We calculated the equivalent dose rate on the surface for all events by using the standard tissue-equivalent ICRU (International Commission on Radiation Units and Measurements) sphere of 30 cm diameter and a density of 1 g cm$^{-3}$ \citep{international1980radiation}. 

\section{Results} \label{sec:results}

\begin{figure}
    \centering
    \includegraphics[width=6in]{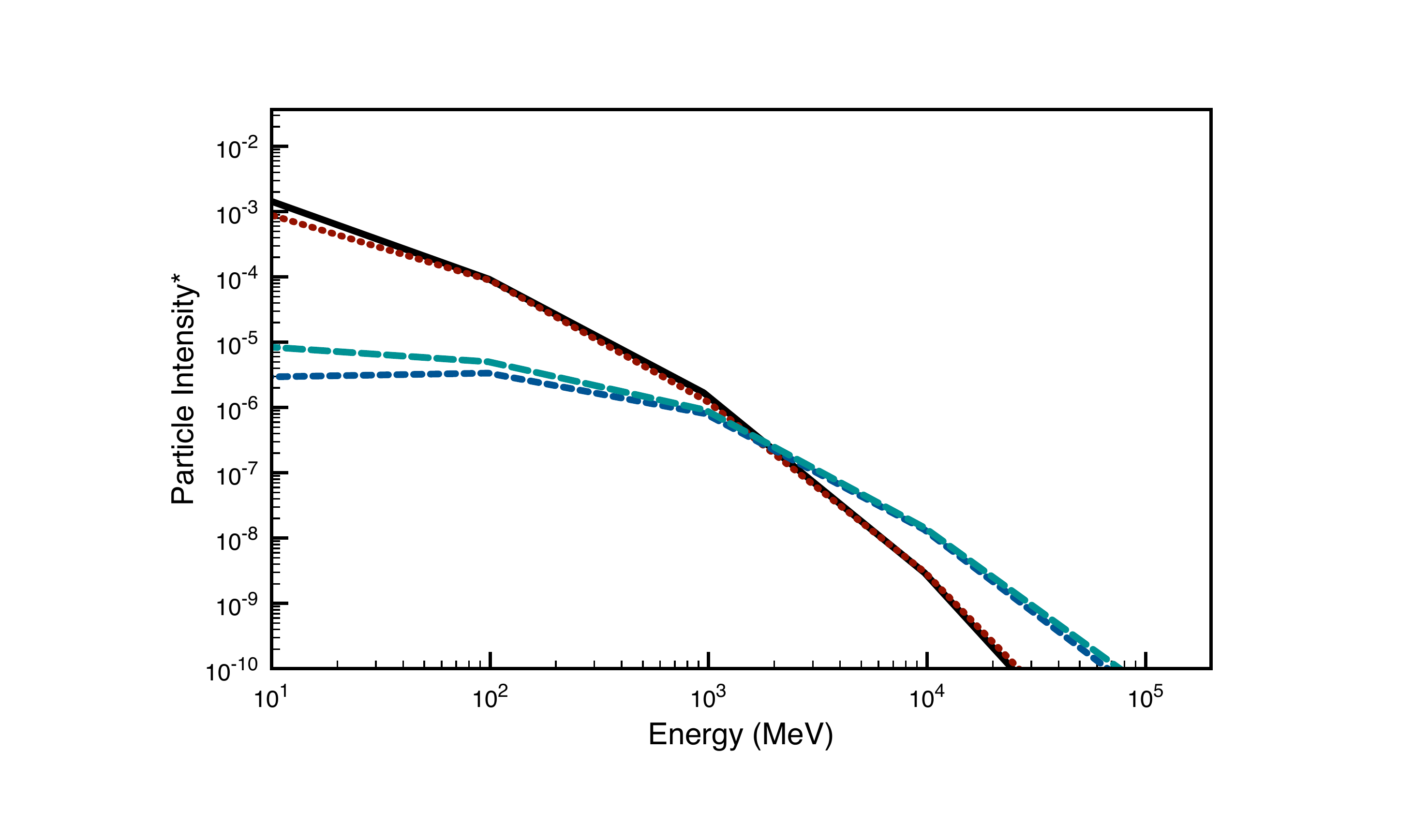}
    \caption{GCR-induced secondary particles on the Martian surface at the Gale Crater modeled using GEANT4. Particle intensity, dI/dE (cm$^{-2}$sr$^{-1}$s$^{-1}$MeV$^{-1}$); e$^{-}$ (solid black), e$^{+}$ (red dotted), $\mu$$^{+}$ (teal dashed), $\mu$$^{-}$ (blue dashed). BON 10 model was used to calculate the incident GCR spectrum \citep{o2010badhwar}.}
    \label{fig:secondaries}
\end{figure}

\begin{figure}
    \centering
    \includegraphics[width=6in]{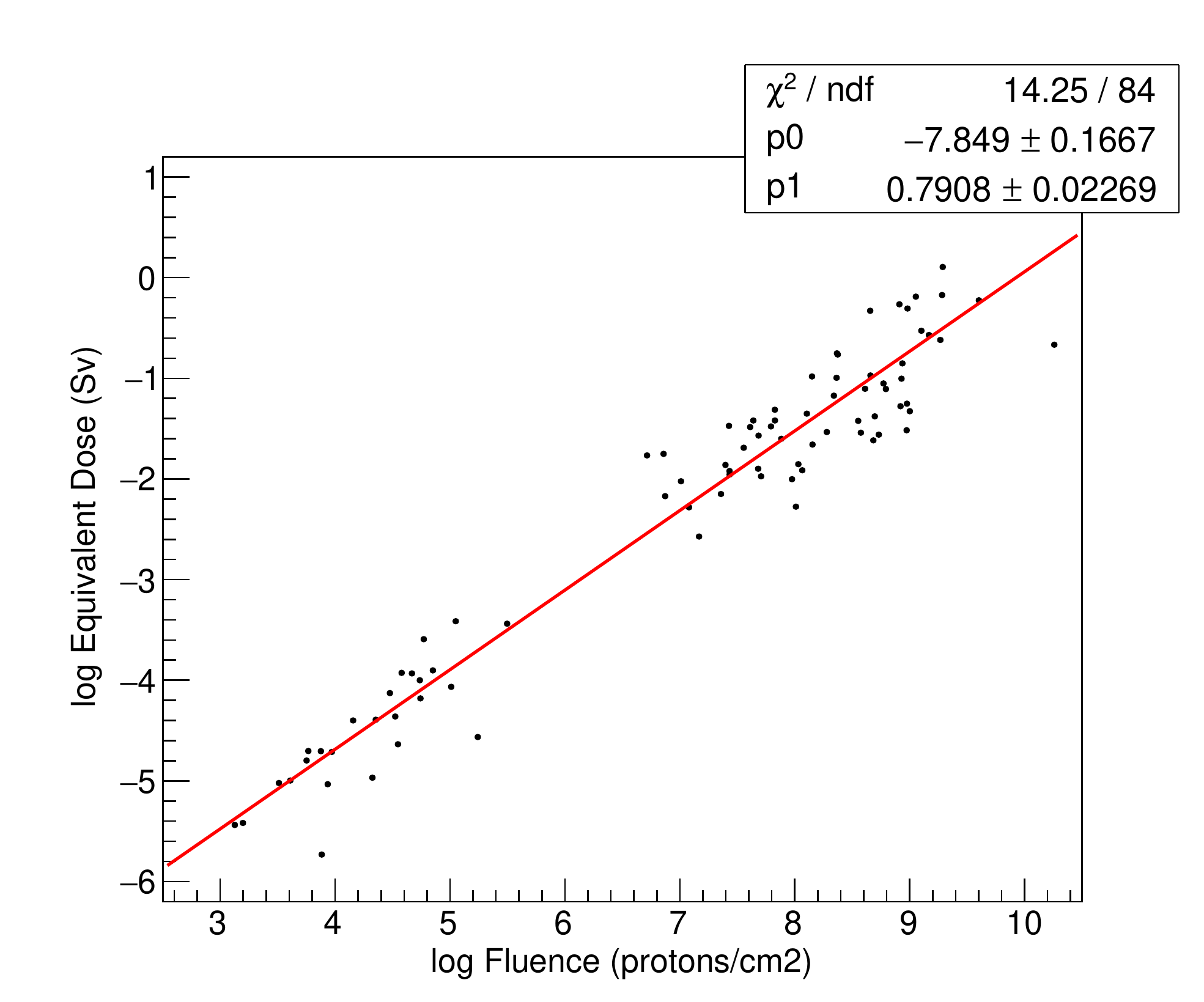}
    \caption{The SPE-induced surface radiation dose from all events as a function of proton fluence on the Martian surface.}
    \label{fig:dosefluence}
\end{figure}

\begin{figure}
    \centering
    \includegraphics[width=6in]{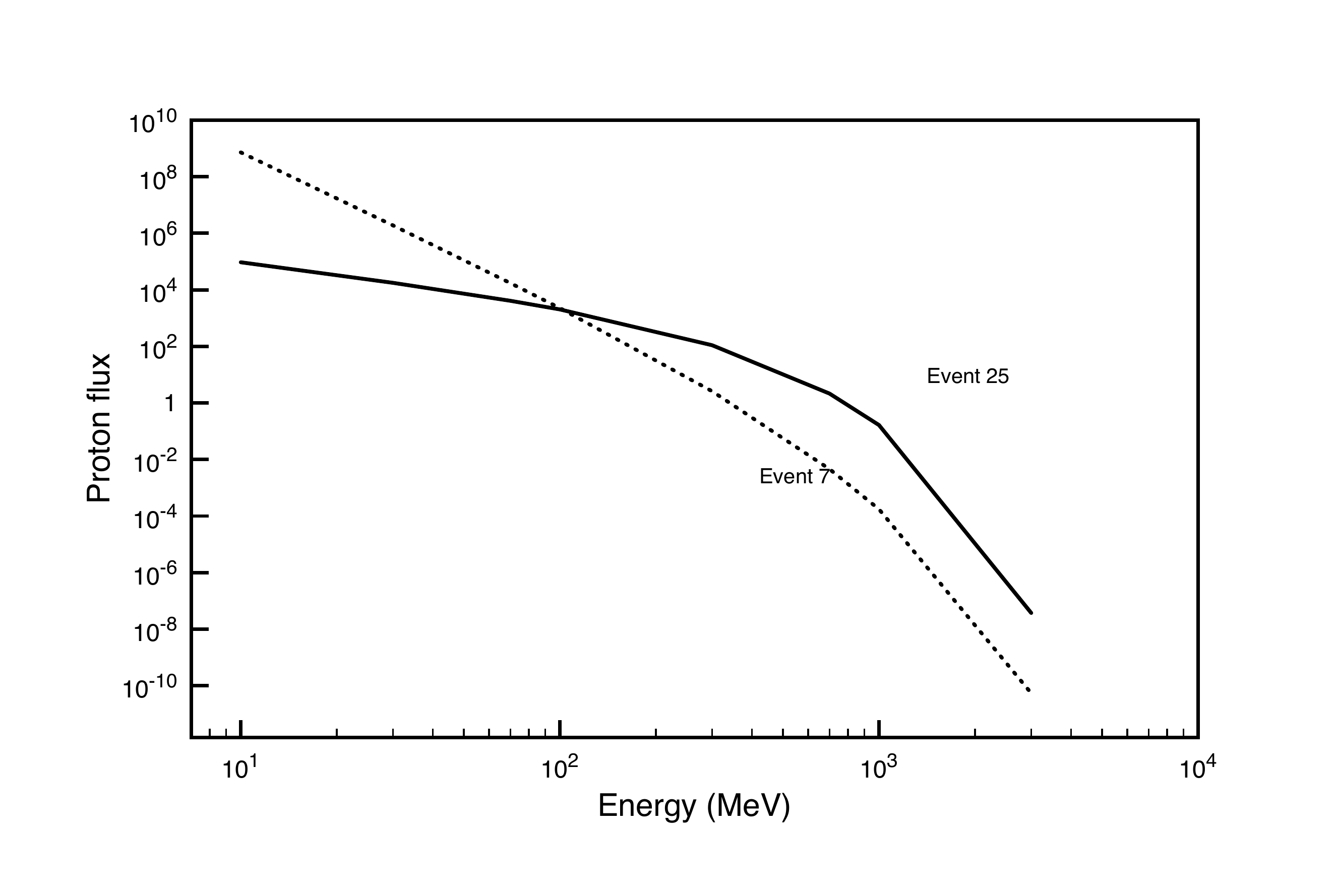}
    \caption{The energy spectrum of events 7 (Jan 23, 2012) and 25 (Sept 1, 2014) from PAMELA \citep{bruno2018solar}.}
    \label{fig:spec1}
\end{figure}

\begin{figure}
    \centering
    \includegraphics[width=6in]{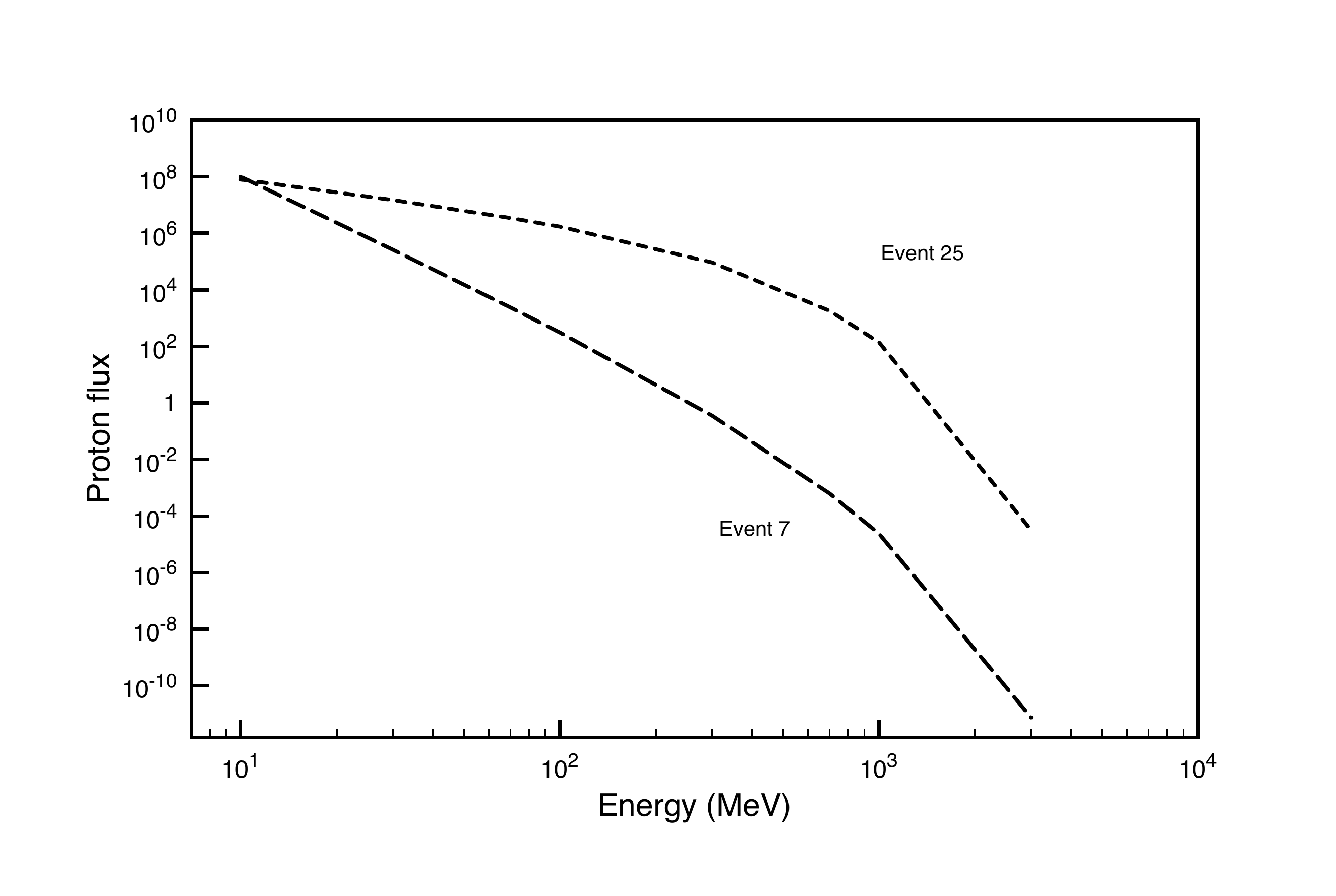}
    \caption{The energy spectrum of events 7 (Jan 23, 2012) and 25 (Sept 1, 2014) normalized to a fluence of 10$^{8}$ protons cm$^{-2}$.}
    \label{fig:spec2}
\end{figure}

\begin{figure}
    \centering
    \includegraphics[width=6in]{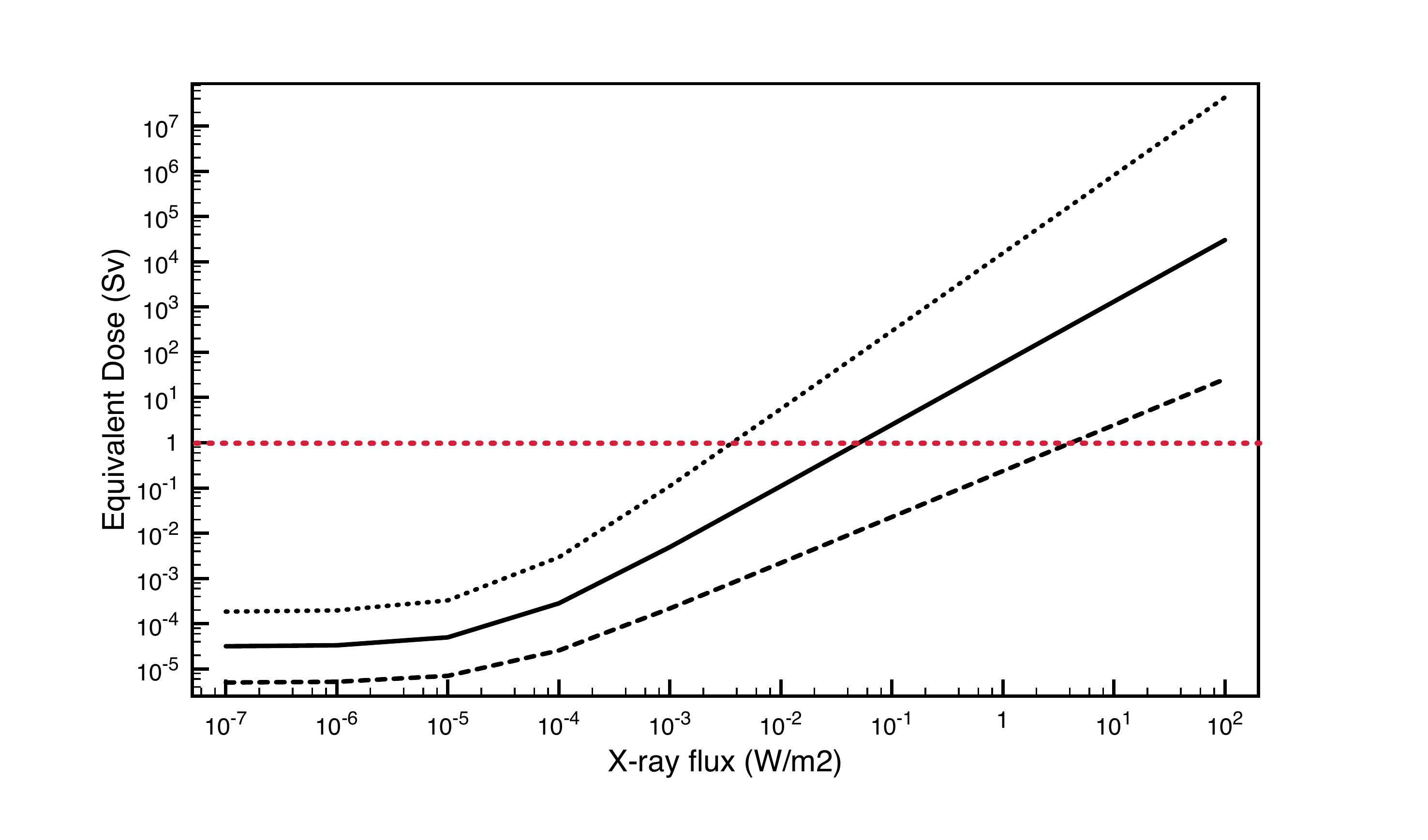}
    \caption{Martian surface equivalent radiation dose corresponding to X-ray flare intensity. The worst case scenario is represented by the dotted line, median scenario by the solid line, and the best case scenario by the dashed line. 1 Sv, which is represented by the red horizontal line is the upper limit of the total equivalent dose considered safe for an astronaut in their entire career.}
    \label{fig:xraydose}
\end{figure}

We first present the simulation results of GCRs interacting with the Martian atmosphere, which we have used for calibration purposes. Then, we show results of SPE-induced radiation dose on the surface from all events. We then demonstrate how radiation dose depends on particle spectrum and then calculate the overall dose as a function of the flare X-ray intensity. Figure \ref{fig:secondaries} shows the energy spectrum of secondary particles on the Martian surface based on the BON10 GCR spectrum \citep{o2010badhwar}, modeled using GEANT4. It can be seen that the lower energy end of particles is dominated by electrons and positrons, and muons for the higher energy end. This is because electrons lose energy easily as opposed to muons because of their small mass. As discussed earlier, we validated our simulations by first computing the background flux of GCRs and comparing them with RAD measurements. We use the energy spectrum of secondary particles to calculate the energy deposition per unit mass in tissue to obtain the radiation dose \citep{matthia2016martian}.  We calculate the equivalent dose in Sieverts by using the ICRU and ICRP (International Commission on Radiological Protection) recommendations \cite{mountford1992recommendations} which assigns different weighing factors to different radiation types. Electrons and photons both have been assigned a factor of 1, whereas for neutrons it ranges from 8 up to 20, depending on their energy. A cumulative exposure to a dose above 1 Sv can cause radiation sickness and induce carcinogenic defects, therefore a dose of 1 Sv is the career limit for a NASA astronaut. The upper limit of radiation exposure for a US worker is 50 mSv annually. The annual background dose from natural sources is $\sim$ 1 mSv and a typical chest X-ray scan is 0.1 mSv \cite{schauer2009ncrp}. Our simulations gave the equivalent dose rate from background GCRs to be 0.59 mSv/day, which is consistent with RAD measurement of 0.64 $\pm$ 0.12 mSv/day \citep{hassler2014mars} within instrumental uncertainties. 

The surface radiation dose for all events as a function of fluence is shown in Figure \ref{fig:dosefluence}. The satellite data has recorded lower fluence because of a limited energy range as compared to terrestrial data, which are estimates between 10 MeV and 10 GeV. Regardless, they follow the same trend which can be clearly seen in the figure. We are able to derive an empirical relation to calculate the surface radiation dose from this plot.
\begin{equation}
    D = F^{p1} \times 10^{p0} 
\end{equation}

Where D is the radiation dose (Sv), F is the particle fluence, p0 = -7.849 $\pm$ 0.1667, and p1 = 0.7908 $\pm$ 0.02269. 
This large variation in the surface radiation dose also reflects the variation in particle acceleration in various events. The radiation dose depends both on the fluence and the spectral shape of particles for a particular event. Higher energy particles contribute more to the surface radiation dose compared to lower energy ones. The energy spectrum of events 7 (Jan 23, 2012) and 25 (Sept 1, 2014) is shown in Figure \ref{fig:spec1}, and the corresponding radiation dose is 8.52$\times$10$^{-7}$ Sv and 7.99$\times$10$^{-6}$ Sv respectively. We normalized the two events to the same fluence of 10$^{8}$ protons cm$^{-2}$, and found the radiation dose to be 4.88$\times$10$^{-4}$ and 1.35$\times$10$^{-2}$ Sv respectively. This difference in radiation dose, which is about a factor of 28, is due to the difference in spectral shapes, which can be clearly seen in Figure \ref{fig:spec2}.

Our results are also consistent with \cite{guo2018generalized}, where for the September 29, 1989 SPE, our dose is 2.21$\times$10$^{4}$ $\mu$Gy day$^{-1}$, compared to 2.12$\times$10$^{4}$ $\mu$Gy day$^{-1}$ of \cite{guo2018generalized}, and for the October 19, 1989 SPE, our dose is 3.20$\times$10$^{3}$ $\mu$Gy day$^{-1}$, compared to 3.13$\times$10$^{3}$ $\mu$Gy day$^{-1}$ of \cite{guo2018generalized}.

In order to estimate the radiation dose from SPEs on long timescales, it is important to incorporate the occurrence rate of major flares on G-type stars. Recent observations from {\it Kepler} and {\it Gaia} has shown that a flare of 5$\times$10$^{34}$ ergs occurs once every 2000-3000 years on Sun-like stars \citep{notsu2019kepler}. Such events are more than 3 orders of magnitude larger than the Carrington event, the largest solar event recorded in history. Although flares are observed in the X-ray and UV bands, no protons have been detected from other stars. In order to solve this issue, we use the relation between the X-ray intensity and particle fluence, which is an empirical relation, derived from extensive flare observations \citep{herbst2019solar}. This empirical relation provides an estimated proton fluence corresponding to photon fluence from a flare.

Radiation dose as a function of X-ray flare intensity is calculated based on equation 4 and the above mentioned empirical relation and is shown in Figure \ref{fig:xraydose}, which has been extrapolated far beyond the observed flare intensity on the Sun. Events towards the higher energy end are highly unlikely in the timescale of the next 100 years but are shown in order to get an estimate of the worst case scenario, which is likely to occur in 100s of Myr timescales. The dotted line corresponds to the worst case scenario, which is a combination of the upper limit of the proton fluence estimate and the dose calculated from equation 4. The solar line is the median and the dashed line corresponds to the best case scenario. The red dashed line corresponds to the equivalent dose of 1 Sv, which is the career limit of astronauts. A Carrington-like event, which was an X45 event, or 4.5$\times$10$^{-3}$ Wm$^{-2}$ on Earth and $\sim$2$\times$10$^{-3}$ Wm$^{-2}$ on Mars, would expose the astronauts to an equivalent dose less than 1 Sv, even in the worst case scenario. In case of a super-Carrington event, or a X450 event, the dose approaches lethal levels in the worst case scenario, however the median case is still safe for astronauts.

We should note that these doses are for individual events and the total radiation exposure would also include background radiation from GCRs throughout the duration of the mission. The equivalent dose rate during the cruise phase is 1.9 mSv day$^{-1}$ and on the Martian surface is 0.7 mSv day$^{-1}$ \cite{hassler2014mars}. During a 180 day cruise to Mars, the equivalent dose exposure would be 340 mSv, and the same during return making it 680 mSv. For a surface mission of 450 and 600 days, the total background dose including transit would be 1.02 to 1.1 Sv respectively \cite{hassler2014mars}. As we have shown, a combination of radiation dose from GCRs and SPEs can be damaging to astronauts, which we describe in the next section. It is therefore crucial to adopt effective shielding strategies in order to counter the adverse impacts on astronaut health.

\section{Discussion and Conclusions} \label{sec:conclusions}
Exposure to SPEs and GCRs is one of the main sources of health risks faced by astronauts in future crewed missions and settlements on Mars. The absence of a significant atmospheric shielding and the lack of a planetary magnetic field provides maximum exposure from energetic charged particles on the Martian surface. Although, the satellite data is more accurate, it has a limited energy range. On the other hand, terrestrial data spans over a broader energy range, but the spectra is based on neutron monitor data, which is less accurate. The terrestrial GLE data spanning over several decades is a useful resource to study the variation in surface radiation dose between events and has aided in better estimating the impact of more energetic events than those available in historical records. The surface radiation dose is governed by two main factors, particle fluence, and spectral hardness. Our calculations are based on a variety of GLEs from both terrestrial and satellite observations over the past several decades, from low-fluence, soft-spectrum events, to high-fluence, hard-spectrum events. 

In worst case scenarios, a Carrington-type event can pose a moderate health risk to astronauts and a more extreme super-Carrington like event can be lethal. The median dose, however, is much smaller for a Carrington-like event and will not pose any noticeable health risk in addition to the risk from exposure to GCRs \cite{schauer2009ncrp}. However, when combined with radiation dose from GCRs, our results show that the surface radiation dose is above the career limit for astronauts (1 Sv) even in the most conservative scenarios. Much of our knowledge about how radiation exposure affects the human body comes from the side effects of radiation therapy. Since radiation therapy does not employ neutron exposure, 1 Sv is approximately equivalent to 1 Gy in patients undergoing treatment. Likewise, since most experimental studies employ low photon energy as their radiation source, 1 Sv is approximately equivalent to 1 Gy in radiobiology publications \citep{Dainiak2003473}. 

Studies of brain cancer patients receiving radiation therapy show that fatigue is a common side effect. The potential occurrence of radiation-induced fatigue is of concern for astronauts who need to be alert when undergoing hazardous space exploration.  Radiation-induced fatigue is characteristic in that it is not typically alleviated by resting. Skin erythema and hair loss are other side effects observed in brain cancer patients receiving radiation therapy \citep{Butler2006517}. HZE particles are a highly-abundant component of galactic cosmic radiation, and have the potential to damage the neurons of the brain causing DNA damage. Neurons have a very slow regeneration rate, allowing DNA damage to accumulate over time. Resulting effects such as premature aging and impaired cognitive ability could impact the exposed astronauts \citep{Craven1994873}.

The skeletal system is also impacted by irradiation, with spaceflight-relevant radiation doses being shown to induce bone loss for as long as four months post-exposure. Bone damage is also a side effect of radiation therapy. Breast cancer patients receiving radiation therapy display rib fracture rates as high as 19\% as a delayed side effect of the treatment \citep{Willey201154}. Women over the age of 65 receiving radiation therapy for cervical cancer had a 65\% increased relative risk of hip fracture \citep{Baxter20052587}. \citep{Kook2015255} carried out an experiment where MC3T3-E1 cells were exposed to radiation doses ranging from 0-8 Gy. The upper limit of this range is comparable to the worst case scenario of radiation exposure from a super-Carrington event. MC3T3-E1 is a cell line of osteoblast precursors from Mus musculus (house mouse). Osteoblasts are cells that function in synthesizing bone. Irradiation of these MC3T3-E1 cells led to an impaired ability to differentiate into mature osteoblasts to carry out bone synthesis, thus a potential contributor to overall bone loss.

A cumulative dose of 1-2 Sv, which we expect from a typical mission would damage white blood cells (WBCs). WBCs are key components of the immune system, are vulnerable to radiation damage due to a high rate of regeneration allowing harmful mutations to display sooner. \cite{Gridley200255} carried out experiments exposing immune cells (WBCs) in mice to irradiation by protons. The immune cells were exposed to radiation doses of 0.5 Gy, 1.5 Gy, and 3.0 Gy in fractions of 1 cGy or 80 cGy per minute. The experiment observed a dose-dependent decline in white blood cell populations, and an impairment on the functionality of the exposed WBCs. This can mediate an impaired immune system with a weakened ability to fight infection, which is a concern for astronauts.

\cite{Delp201629901} investigated the mortality rate of radiation-induced cardiovascular disease (RICVD) in three different groups of astronauts: astronauts who had flown in low-Earth orbit, astronauts who had not ever flown in low-Earth orbit, and the Apollo lunar astronauts. The Apollo lunar astronauts are the only humans to have traveled outside of the protection of the Earth’s magnetic field, and showed a cardiovascular disease mortality rate that was 4-5 times higher than the other astronaut groups. This reinforced the increased danger of deep space radiation exposure, in comparison to that experienced within the protection of the geomagnetic field. Cardiovascular disease is the leading cause of death worldwide. Studies of the Chernobyl disaster survivors have shown that a radiation dose exposure as low as 0.15 Gy can increase the risk of developing radiation-induced cardiovascular disease. Similarly, heart disease and hypertension have been observed as the two most common late side effects of radiation exposure from the survivors of the atomic bombings of Hiroshima and Nagasaki in 1945 \citep{Hughson2018167}.

A late side effect of radiotherapy for non-small cell lung cancer is the development of radiation-induced lung fibrosis. Fibrosis is a pathological process where injured tissue being repaired is replaced by scar tissue rather than functional, healthy tissue. Radiation-induced lung fibrosis can result in clinical symptoms such as respiratory insufficiency, dry cough, and dyspnea (shortness of breath) \citep{Ding20131347}. 

Studies on the male and female reproductive systems have shown that a radiation dose exposure of over 0.35 Gy to the male testis can result in aspermia (lack of semen in an ejaculation). Permanent aspermia can result at a radiation dose exposure of over 2 Gy. For females, irradiation of the ovaries with a radiation dose of 4 Gy is observed to produce 30\% sterility incidence in women under 40 years of age, and 100\% sterility incidence in women over 40 years of age \citep{Oglivy-Stuart1993109}. 

Acute radiation syndrome (ARS) is a severe short-term illness caused by radiation exposure. The Centers for Disease Control and Prevention (CDC) classify 0.7 Gy as the threshold of radiation exposure for ARS to occur, but report that mild symptoms are observable at doses as low as 0.3 Gy \citep{Jones202039}. ARS presents in three medically-distinct presentations: the hematopoietic syndrome, the gastrointestinal syndrome, and the neurovascular syndrome. The hematopoietic syndrome presents at 0.3-5 Gy radiation doses and therefore is the most relevant for radiation exposure levels expected for astronauts visiting Mars. The hematopoietic syndrome impacts the formation of red and white blood cells, and can have consequences such as increased susceptibility to infection, and anemia.

The increased cancer risk that astronauts may experience in response to radiation exposure is a notable concern. \cite{Hall200383} report that survivors of atomic bombings show increased incidence of carcinoma developments in various tissues in a radiation dose-dependent relationship up to around 2.5 Sv. NASA has radiation limits set for astronauts so that the upper bound of acceptable risk is a 3\% increased cancer mortality rate due to radiation exposure. \cite{Cucinotta20149} carried out analysis of the cancer risk in astronauts in the International Space Station (i.e. in low Earth orbit) and found that female astronauts in low-Earth orbit may exceed NASA’s radiation limit within 18 months, and male astronauts within 24 months. \cite{Straume2015259} also discussed how the radiation limit would be exceeded in a mission to Mars. HZE particles, which are abundant in deep space radiation, have been shown to induce tumors of higher lethality in comparison to Earth-based gamma rays \citep{Cucinotta20149}. Along with the threat posed by SPEs, an increase in cancer risk can also result from the occurrence of ultraviolet radiation-induced inflammation. Experiments in laboratory mice showed that inflammation as a result of ultraviolet irradiation produced increased angiogenesis (the formation of new blood vessels) and increased metastasis (the spread of cancer around the body) \citep{Bald2014109}.  

{\it Kepler}, {\it Gaia} and {\it TESS} observations of G-type stars has provided us with the occurrence rate of energetic events over long timescales \cite{notsu2019kepler}. Based on these observations it can be estimated that the likelihood of a Carrington-like events directed towards Mars is very small for a 3-year mission to Mars. However, such an event is highly likely on longer, 50-100-year timescales, relevant for human settlements on the planet. Super-Carrington events are likely to occur on 1000-year timescales and are less probably in either of the above scenarios. Our estimates can be improved with a better understanding of the relationship between flare energy and particle fluence at higher energies, and better particle transport models in the interplanetary medium. 

\acknowledgments

This work is supported by the NYUAD Institute research grant G1502. We thank members of the GEANT4 collaboration who developed the software (\url{geant4.cern.ch}) used for this work. This research was carried out on the High Performance Computing resources at New York University Abu Dhabi.



\bibliography{hspe_mars}
\end{document}